\documentclass[fleqn,usenatbib]{mnras}

\usepackage{xcolor}

\usepackage{newtxtext,newtxmath}

\usepackage[T1]{fontenc}

\DeclareRobustCommand{\VAN}[3]{#2}
\let\VANthebibliography\thebibliography
\def\thebibliography{\DeclareRobustCommand{\VAN}[3]{##3}\VANthebibliography}

\usepackage{graphicx}	
\usepackage{amsmath}	
\usepackage{subcaption}
\usepackage{booktabs}
\usepackage{enumitem}

\usepackage{scalerel}
\usepackage{tikz}
\usetikzlibrary{svg.path}
\definecolor{orcidlogocol}{HTML}{A6CE39}
\tikzset{
  orcidlogo/.pic={
    \fill[orcidlogocol] svg{M256,128c0,70.7-57.3,128-128,128C57.3,256,0,198.7,0,128C0,57.3,57.3,0,128,0C198.7,0,256,57.3,256,128z};
    \fill[white] svg{M86.3,186.2H70.9V79.1h15.4v48.4V186.2z}
                 svg{M108.9,79.1h41.6c39.6,0,57,28.3,57,53.6c0,27.5-21.5,53.6-56.8,53.6h-41.8V79.1z M124.3,172.4h24.5c34.9,0,42.9-26.5,42.9-39.7c0-21.5-13.7-39.7-43.7-39.7h-23.7V172.4z}
                 svg{M88.7,56.8c0,5.5-4.5,10.1-10.1,10.1c-5.6,0-10.1-4.6-10.1-10.1c0-5.6,4.5-10.1,10.1-10.1C84.2,46.7,88.7,51.3,88.7,56.8z};
  }
}

\newcommand\orcidicon[1]{\href{https://orcid.org/#1}{\mbox{\scalerel*{
\begin{tikzpicture}[xscale=1,yscale=-1, transform shape]
\pic{orcidlogo};
\end{tikzpicture}
}{|}}}}

\title[Outbursts in IPs]{A Systematic Search for Optical Outbursts in IPs Using ASAS-SN}

\author[J. M. Mendelsohn et al.]{
J. M. Mendelsohn,$^{\orcidicon{0009-0006-8357-000X}}$$^{1,2}$ \thanks{E-mail: jake.mendelsohn@physics.ox.ac.uk}
S. Scaringi,$^{\orcidicon{0000-0001-5387-7189}}$$^{1,3}$
M. Veresvarska,$^{\orcidicon{0000-0002-0146-3096}}$$^{1}$
K. Iłkiewicz, $^{\orcidicon{0000-0002-4005-5095}}$$^{4}$
\\
$^{1}$Centre for Extragalactic Astronomy, Department of Physics, Durham University, South Road, Durham, DH1 3LE, UK\\
$^{2}$Astrophysics, University of Oxford, Denys Wilkinson Building, Keble Road, Oxford, OX1 3RH, UK\\
$^{3}$INAF -- Osservatorio Astronomico di Capodimonte, Salita Moiariello 16, I-80131 Naples, Italy\\\
$^{4}$Nicolaus Copernicus Astronomical Center, Polish Academy of Sciences, Bartycka 18, 00-716 Warsaw, Poland
}

\date{Accepted 2026 April 28. Received 2026 April 20; in original form 2025 November 21}

\pubyear{\the\year{}}

\begin{document}
\label{firstpage}
\pagerange{\pageref{firstpage}--\pageref{lastpage}}
\maketitle

\begin{abstract}
Cataclysmic variables can show rapid increases in optical flux. Intermediate polars (IPs), a subset with strong magnetic fields that disrupt the inner accretion disc, have been thought to possess truncated discs that rarely undergo the disc-instability outbursts seen in dwarf novae. However, the discovery of micronovae and magnetic-gating bursts suggests that such events may occur even without a fully developed disc. Using data from the All-Sky Automated Survey for Supernovae (ASAS-SN), we identify a previously unrecognised population of short-timescale optical outbursts in IPs. Initial energy estimates indicate that at least half of these bursts may be consistent with micronovae, though limited cadence reduces our ability to classify each event with high confidence. These detections should therefore be regarded as evidence of short outbursts in IPs rather than definitive micronova identifications. Our results show that such bursts are more common in IPs than assumed and may include a substantial fraction of micronovae. Simulations reveal that if micronovae occur once per year at regular intervals, up to 30\% of the shortest bursts could be missed over a 10-year observing baseline. Under the same assumptions, and assuming all IPs display micronovae, we would expect 50--70\% of IPs to show these bursts, yet only 14\% of known IPs in our sample do so. This discrepancy suggests that not all IPs undergo micronovae. Overall, this work establishes the first comprehensive set of short burst detections in IPs, providing a foundation for future investigations and a list of candidate micronova systems for follow-up analysis.

\end{abstract}




\section{Introduction}

Cataclysmic variables (CVs) are binary systems in which a white dwarf (WD) accretes from a main sequence companion. Accretion occurs when the donor star overfills its Roche lobe, forming an accretion disc around the white dwarf \citep{hellier2001,warner2003cataclysmic}.

CVs are broadly classified into two categories based on the magnetic field strength of the WD. 'Non-magnetic' CVs are those where the WD's magnetic field is too weak to disrupt the accretion disc ($\lesssim10^6$ G), whereas 'magnetic' CVs possess a white dwarf with a strong enough magnetic field ($\gtrsim10^6$ G) to partially or entirely inhibit the formation of an accretion disc. 'Magnetic' CVs can be further categorised depending on the extent to which they are able to disrupt the accretion disc and the synchronicity of the spin and orbital periods. The systems with the strongest magnetic fields ($\gtrsim10^7$ G) are able to prevent the accretion disc from forming entirely. These systems are known as polars, with the name being derived from strongly polarised emission observed at the magnetic poles of the WD.  In these systems, the WD’s intense magnetic field channels accreting material directly onto its poles, where cyclotron emission produces a distinctive linear and circular polarisation. In the majority of polars, the spin of WD is synchronised with the orbit of the binary.  

There exists a separate class of 'magnetic' CVs that have an intermediate magnetic field ($10^6-10^7$ G) which usually only disrupt part of the inner accretion disc. These systems are known aptly as intermediate polars (IPs). In IPs, the inner radius of the accretion disc is truncated at the Alfv\'en radius, within which material will flow along magnetic field lines onto the poles of the WD. The existence of a disc in IPs is governed by both the circularisation radius, $R_{\mathrm{circ}}$, and the Alfven radius, $R_A$. The circularisation radius refers to the radius at which material from the donor star settles into a Keplerian orbit around the WD. If $R_A>R_{\mathrm{circ}}$, the stream of matter from the donor is captured by the field before a disc is able to form and material latches onto the magnetic field lines and accretes onto the magnetic poles via accretion streams or curtains \citep{hellier2001}. More specifically, disc truncation will occur at the magnetospheric radius, $r_M = \Lambda R_A$, due to the misalignment of the magnetic field with respect to the disc causing the ram pressure to make a dent in the magnetosphere \citep{Ghosh_1979,M_nkk_nen_2022}. $\Lambda$ is a numerical factor usually assumed to be 0.5 in neutron stars \citep{M_nkk_nen_2022}. In IPs, the WD's spin is not synchronised with orbital period and we observe a periodic modulation on the WD spin period due to emission within the accretion column on the WD pole. In rare cases, the magnetic field of IPs can be strong enough to inhibit the entire accretion disc, forming disc-less IPs such as V2400 Oph \citep{Buckley_1995}.

CVs experience short rises in brightness which are generally known as outbursts, and come several different forms. The most commonly occurring and well understood bursts in CVs are dwarf novae, which occur due to a sudden increase in mass accretion rate as a result of thermal viscous instabilities within the accretion disc \citep{Smak_1984,Osaki_1996,Lasota_2001}. The size of the accretion disc directly impacts the duration of these bursts, with the longest bursts being found at longer orbital periods \citep{Lasota_2001,Hameury2017}. Specifically, these bursts can range from a few days to weeks, with recurrence timescales of days to decades. Systems that experience dwarf novae are also susceptible to superoutbursts. These outbursts are 0.7 mag brighter than typical dwarf novae and last 5 to 10 times longer than the normal ones \citep{Buat_M_nard_2002}.

Since magnetic CVs often have truncated accretion discs, they are able to maintain stability in "cold equilibrium", when material in the disc is not fully ionised. However, it remains uncertain whether the size of the accretion disc truly suppresses dwarf novae or merely diminishes their amplitude. In magnetic systems, two other types of outburst appear as possible alternatives. The first of these are magnetically gated bursts, where material builds up at the magnetospheric radius and is prevented from accreting onto the WD by spinning material \citep{Littlefield2022,Scaringi2017}. Accretion is halted by the WD magnetic field until enough pressure is built up, to release a short burst of accretion. These bursts have extremely short durations ($\leq6$ hours) and recur on timescales of a few days \citep{Littlefield2022}, and sometimes even shorter \citep{Scaringi2017}.

Micronovae are another proposed type of burst that are recently being observed in an increasing number of systems, with multi-peaked turbulent outburst shapes similar to Type 1 X-ray bursts in accreting neutron stars \citep{Scaringi_2022b}. These bursts have short durations, lasting from several hours to a few days, but cannot be attributed to magnetic gating bursts due to the significant energies released ($10^{38}-10^{40}$ erg). The currently hypothesised origin of these bursts is localised thermonuclear reactions on the magnetic poles of the WD, however this is yet to be confirmed through observation \citep{Scaringi2022}. Until now, only six targets have been reported to show these bursts \citep{Scaringi_2022b,Veresvarska_2024,veresvarska2025,Ilkiewicz2024,Irving_2024, veresvarska2025}. One of these targets, TV Col, was reported to display P Cygni profiles with outflow velocities of 3870 kms$^{-1}$ \citep{Szkody_1984}. This implies that the material is being expelled rather than accreted, contrary to what occurs in magnetic gating. With the advent of new all-sky surveys such as BlackGEM \citep{groot2024blackgem} and LSST \citep{ivezic2019lsst} many more are expected to be found, substantially increasing the current sample. The three types of bursts discussed here have been classified previously in diagnostic diagrams developed by \citet{Ilkiewicz2024} and these will be applied within this work.

Micronovae are thought to be a result of interactions within the accretion column above the magnetic poles of the WD and therefore do not require an accretion disc to occur. As a result, it is possible that these outbursts are widespread throughout the population of IPs and have been missed previously due to their short duration ($\sim$12 hours-2 days). Furthermore, \citet{Hameury2017} find that the fraction of outbursting systems in non-magnetic CVs is 42\%. Assuming a similar ratio in IPs, they expected a 16 IPs to show dwarf novae whereas only 5 were found. Motivated by these facts, we explore the broader population of IPs for signatures of outburst activity.  To do this, we perform a systematic search for outbursts within the All Sky Automated Survey for Supernovae (\textit{ASAS-SN}) \citep{Shappee_2014,Kochanek_2017,Jayasinghe_2019} and also the Zwicky Transient Facility (\textit{ZTF}) \citep{Masci_2018}. For each burst found, we carried out follow-up analysis using data from the Transiting Exoplanet Survey Satellite (\textit{TESS}) to search for counter-parts or further evidence of bursts. We present the results of this analysis in Section \ref{sec:BurstAnalysis} and use similar techniques as \citet{Ilkiewicz2024} to compare bursts in Section \ref{sec:BurstComparison}.

For each detected outburst, we attempt to calculate energy, duration, and peak luminosity using distances derived from GAIA EDR3 \citep{Bailer_Jones_2021} since these metrics allow us to differentiate between outburst types. In order to classify our results, we follow the methodology of \citet{Ilkiewicz2024}. However, because of the sparcity of \textit{ASAS-SN} sampling, this was erroneous and we attempt to quantify this error in Section \ref{sec:ErrorSim}. Finally, we simulate how the proportion of expected micronovae detected changes with observing time using our method and sampling. The results of these simulations are presented in Section \ref{sec:DetectionRate}.     

\section{Data and Observations}
In this section we present and summarise the details of the sample (Section \ref{sample}) and the ground based (Sections \ref{sec:ASAS-SN} and \ref{sec:ZTF}) and space based (Section \ref{sec:TESS}) data used in this work.

\subsection{IP Sample}
\label{sample}

The population of IPs used in this study was compiled by Koji Mukai and last updated in 2021, containing 202 candidate IPs \footnote{https://asd.gsfc.nasa.gov/Koji.Mukai/iphome/catalog/alpha.html} \citep{Mukai_2023}. Within this list, individual IPs are classified according to a star classification scheme, with 1* IPs being considered 'doubtful' candidates and 5* IPs being considered 'ironclad' IPs. While being a few years out of date, Mukai's list was chosen as it remains the most robust list of confirmed IPs, so we avoid selecting misidentified systems. Within this study, we only examine 4* and 5* IPs. The difference between these classifications being that there is little uncertainty on the spin period in ironclad systems. Therefore, in this study we examine the long-term light curves of 71 confirmed and ironclad IPs, which we collectively refer to as ironclad IPs.

\subsection{ASAS-SN}
\label{sec:ASAS-SN}
For each target, we search a combination of \textit{ASAS-SN} g-band and V-band data. The g-band is centred at 475 nm with a width of 140 nm while the V-band is centred at 550 nm with a width of 90 nm. This data was downloaded directly from the \textit{ASAS-SN} website \footnote{https://asas-sn.osu.edu/photometry}. ASAS-SN was chosen as the main data source due to its long-term coverage (10+ years) and it having light curves for all 71 of the confirmed IPs in our population. We convert between \textit{ASAS-SN} flux and luminosity by applying the relation, 
\begin{equation}
    L[\text{erg s}^{-1}] = \frac{4\pi F_{ASAS-SN}[Jy]\times \nu[Hz]\times d^2[cm]}{10^{23}},
\label{eq:lumi}
\end{equation}
where $\nu$ is the frequency at band centre and $d$ is the distance to the source. All distances are taken as the median value of GAIA parallaxes \citep{Bailer_Jones_2021}.

\subsection{TESS}
\label{sec:TESS}
During the course of this work, \textit{TESS}, observed the IP IGR J04571+4527 (TCID: 65820714) in sector 86 between November 21$^{\mathrm{st}}$ and December 18$^{\mathrm{th}}$, 2024. \textit{TESS} has a bandpass range of 600-1000nm and has the capability to measure 20s or 2 minute cadence. This data was downloaded from the Barbara A. Misulski Archive for Space Telescopes (MAST) and we use the Simple Aperture Photometry (SAP) data to retain more of the original signal and preserve intrinsic variability \footnote{https://mast.stsci.edu/portal/Mashup/Clients/Mast/Portal.html}. \textit{TESS} data was downloaded from MAST using the \texttt{LightKurve} package in python \citep{lightkurve}. Furthermore, for simulations in Section \ref{sec:ErrorSim}, \textit{TESS} sector 38 of ASASSN-19bh was used in simulations. For TV Col in Section \ref{sec:DetectionRate}, sector 32 between BJD-2457000 2197 and 2200 days was used as the example burst as shown in \citet{Scaringi_2022b}. 

\subsection{ZTF}
\label{sec:ZTF}

Archival data from the Zwicky Transient Facility (ZTF) was also used to support this work. ZTF is a time-domain survey ran on the Samuel Oschin  telescope at the Palomar Observatory. This survey operates in three optical bands such as the g-band $(\sim4087-5222$ \AA), the r-band $(\sim5600-7316$ \AA), and the i-band $(\sim6883-9009)$ \AA. Where possible, ZTF data for each ironclad IP was obtained over a shorter archival period of $\sim7$ years. Of the 202 IP candidates, only 75 had available ZTF data of sufficient quality as a result of only covering the northern hemisphere. This data was compiled by Martina Veresvarska and obtained directly from the NASA/IPAC Infrared Science Archive \footnote{https://irsa.ipac.caltech.edu/Missions/ztf.html}. 

\section{Methods}

\label{sec:BIA}

We initially apply a simple algorithm to determine the fraction of confirmed and ironclad IPs, as described in Section \ref{sample}, showing burst-like behaviour. This algorithm is initially applied to \textit{ASAS-SN} data and returns the detection of a burst if three consecutive data points, $3\sigma$ above the quiescent average, are present. The quiescent average is calculated through an iterative $\sigma$-clipping procedure. We first apply a $3\sigma$ clipping algorithm to the light curve to remove outliers associated with burst events. Then the quiescent average is determined by taking a rolling average with a window of 50 data points. Following this, the remaining IPs are examined by eye to find any other outbursts that are missed. Missed outbursts are a result of less than three consecutive data points being detected above threshold or in some cases, the three data points are split by a detection that falls below threshold. Due to the low sample size, it was possible to examine each system that was not identified by the algorithm, looking for significant detections above the baseline. The above algorithm is also applied to \textit{ZTF} data however this data set only contains 31 out of the 71 total confirmed and ironclad IPs. As a result, we mainly discuss the use of \textit{ASAS-SN} in this work.

In line with \citet{Ilkiewicz2024}, we characterise each burst by its peak luminosity, total optical energy, and duration. We choose to omit the burst frequency metric because the seasonal gaps in ASAS‑SN coverage can bias its determination. Accurately determining each bursts duration, integrated energy, and peak luminosity is challenging due to the low cadence of \textit{ASAS-SN} and lack of full burst coverage in most cases. Total optical energy is calculated by integrating between the burst and the quiescent average across its entire duration. This is done using simple trapezoidal integration due to the low number of data points. Due to the low cadence of \textit{ASAS-SN}, many bursts are often not accurately mapped and had missing data gaps between the start and end point of the burst. To navigate this, upper and lower limits are used, with the final value being quoted as a range between the lower and upper limits. The lower limit only included the data points above the $3\sigma$ threshold whereas the upper limit included these points and the consecutive points before and after the burst. Finally, the peak luminosity is simply taken as the luminosity of the maximum point detected in the burst, where luminosity in erg s$^{-1}$ is calculated according to equation \ref{eq:lumi}. For sources exhibiting multiple bursts, the reported peak luminosity is given by the mean of the peak luminosities from the best sampled bursts in the system. For sources with a single burst, the error on this value is dominated by the error on the distance to the object, given by the GAIA parallax \citep{Bailer_Jones_2021}. The formula to determine the error in this case is described in equation \ref{eq:err_single}. For multiple bursts, the error is often dominated by the dispersion of the peak luminosities and is propagated according to equation \ref{eq:err_multi}.
 
\section{Results}
In Section \ref{sec:BurstAnalysis} we report all of the bursts found using \textit{ASAS-SN}. Following from this, in Section \ref{sec:BurstComparison} we compare bursts using diagnostic diagrams developed by \citet{Ilkiewicz2024}.

\begin{table*}
    \centering
    \setlength{\tabcolsep}{4pt}
    \begin{tabular}{cccccccccc} 
        \toprule
        \textbf{Name} & \textbf{Class} & Number & \textbf{Peak Luminosity ($10^{34}$erg/s)} & \textbf{Total Energy ($10^{38}$erg)} & \textbf{Duration (d)} & \textbf{$\bar{N}$} & $P_{\mathrm{orb}}$ (hrs) & \textbf{s ($10^{34}$erg/s)}\\ 
        \midrule
         Ex Hya  & DNe*  & 2 & $0.024\pm0.002$ & 0.194-0.233 & 2.94-7.22 & 8.0 & 1.64 & 0.0028\\
         HT Cam  & DNe* & 2 & $0.24\pm0.2$ & 0.186-22.5 & 1.52-40.00 & 6.7 & 1.43 & 0.25\\
         IGR J17014-4306 & MNe   & 6 & $0.9\pm0.4$ & 1.91-35.4 & 0.805-7.43 & 7.2 & 12.8 & 0.57\\
         NY Lup & MNe & 4 &$4.0\pm0.5$  &  & 0.871-16.32 & 4.25 & 9.87 & 0.53\\
         V1025 Cen   & DNe  & 2 & $0.05\pm0.01$   & 0.121-0.573 & 1.00-5.52 & 8.5 & 1.41 & 0.019\\
         V1460 Her & DNe*  & 2 & $0.10\pm0.02$  & 4.44-7.00 &9.42-17.5 & 14.0 & 4.99 & 0.023\\
         V2731 Oph & MNe  &  1 &   $21.0\pm5.0$      & 6.21-79.9 &0.0650-2.01 & 6.0 & 15.4 & -\\
         AX J1853.3-0128 & DNe   & 1 &  $0.081\pm0.009$    & 0.473-0.810 &1.05-2.73 & 7.0 & 1.45 & -\\
         RX J2015.6+3711  & MNe & 1 &$1.0\pm0.3$ & 7.35-23.9 & 1.02-11.1 & 3.0 & 12.8 & -\\
         IGR J04571+4527& - & 1 & -& - & - & - & - & -\\
         IGR J18173-2509& MNe & 2 &$11.0\pm9.0$ &0.0554-100 & 0.00127-6.98 & 2.0 & 1.53 & 0.95\\
         V1223 Sgr & MNe & 1 & $1.95\pm0.07$ & 0.0278-64.9& 0.00240-12.0 & 3.0 & 3.37 & -\\
         DO Dra & DNe* & 2 & $0.07\pm0.03$& 1.26-2.64 & 8.04-13.56 & 10.0 & 3.97 & 0.049\\
         IGR J19267+1325 & MNe & 1 & $1.4\pm0.5$ &5.40-36.6 & 0.996-10.8 & 4.0 & 3.45 & -\\
         EI Uma & MNe* &  2 & $2.6\pm0.7$ &43.0-77.2 & 5.02-9.62 & 13.5 & 6.43 & 0.74\\
         V598 Peg & DNe & 1 &$0.020\pm0.007$ &0.187-0.441 & 3.96-9.94 & 3.0 & 1.39 & -\\
         CTCV J2056-3014 & DNe* & 12 &$0.10\pm0.06$ & 3.63-6.03& 4.78-10.7 & 12.2 & 1.76 & 0.051\\
         CC Scl & DNe* & 14 & $0.06\pm0.01$ & 1.36-2.37 & 4.17-11.49 & 12.2 & 1.41 & 0.026\\
         TV Col & Mne* & 24 & $0.582\pm0.1$ & 2.70-7.28 & 1.67-3.67 & 9.7 & 5.49 & 0.25\\
         DW Cnc & Mne* & 1 & $0.21\pm0.01$ & 0.0037-4.6 & 0.0024-5.98 & 3 & 1.44 & -\\
        \bottomrule
    \end{tabular}
    \caption{Summary of all detected outbursts in ironclad IPs, listing their intrinsic properties. The columns show the source name, preliminary burst classification, number of bursts, peak luminosity, total energy, duration, average number of data points per burst, the orbital period of the system, and the standard deviation of the peak luminosities. MNe = Micronova; DNe = Dwarf nova; * = Classification matches that from literature \citep{Hameury2017,Ilkiewicz2024, Qian_2017,Hameury2022, Szkody_2002, veresvarska2025}. Total energy and duration are represented by ranges between lower and upper bounds, denoted by a hyphen. All values are averages across all bursts.}
    \label{tab:outbursts}
\end{table*}

\subsection{Burst Search and Analysis}

\label{sec:BurstAnalysis}

Using the search algorithm described in Section \ref{sec:BIA}, 14 IPs showing at least 1 burst within the 10 year archival \textit{ASAS-SN} observing period are found. A further 4 IPs showing outbursts are found by eye following the initial search and only a single burst in V598 Peg is found within the ZTF data. On inspection, the four IPs missed in the initial run of the search algorithm each showed two clear detections above threshold and additional points below it, but still formed a clear rising or falling structure—or displayed interrupted burst profiles—with one point between three that briefly dropped below threshold. The burst in V598 Peg is observed in both the r and g bands at $\sim 1175$ JD-2458000. During this study, another bursting IP, IGR J04571+4527, is found within sector 86 of \textit{TESS} data and is also included in Table \ref{tab:outbursts} but properties can not be calculated due to a lack of simultaneous \textit{ASAS-SN} data, required to calibrate. This burst was found during a by-eye examination of newly released \textit{TESS} sector data and is included, separate to the main body of this work, as another candidate and its lightcurve is shown in Figure \ref{fig:appendTESS}. Table \ref{tab:outbursts} summarises the measured properties for all 19 bursts detected by \textit{ASAS‑SN} and \textit{ZTF}, and gives the number of bursts observed over a 10 year span for \textit{ASAS-SN} and 7 year span for \textit{ZTF}. Each quantity in the table is the mean value computed across all bursts seen in the system, so individual events may deviate from these averages. We note that the lower bounds on the burst durations reported in Table \ref{properties} are, in some cases, small and unphysical. This arises from the observing strategy of \textit{ASAS-SN}, which takes three consecutive measurements within a short nightly window. As a result, the lower limits on duration can be underestimated, and should therefore be interpreted with caution. We also note that we do not attempt to disentangle systems that might exhibit both micronovae and dwarf novae; if such mixed behaviour occurs, it could bias some of the mean values reported here. In the case of systems with >10 bursts, we aim to take the two most well covered bursts in the V and g bands as we find that in some cases the data is unreasonably skewed by including more uncertain bursts when a sample of well covered bursts is available. These bursts ideally have the most number of data points and lack a large time gap on either side of the burst. For CC Scl, these bursts begin at $\sim$BJD-2457000 -150, 1080, 1290, 1460, 2190, 3600. For CTCV J2056-3014 these are only in the g-band and located at $\sim$BJD-2457000 1350, 2090, 2430, 2530, 3450, 3600. Finally, for TV Col, these bursts are in in the V-band and g-band and are located at $\sim$BJD-2457000 1020, 1103, 1880, 2250, 3015, 3657.

\subsection{Burst Comparison and Preliminary Classification}
\label{sec:BurstComparison}

\begin{figure*}
    \centering
    
    \begin{subfigure}[t]{0.48\textwidth}
        \centering
        \includegraphics[width=\linewidth]{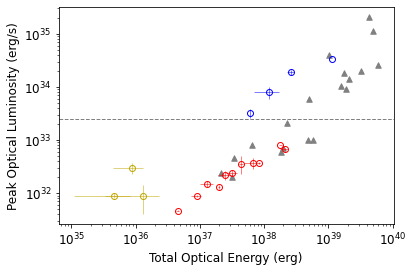}
    \end{subfigure}
    \hfill
    \begin{subfigure}[t]{0.48\textwidth}
        \centering
        \includegraphics[width=\linewidth]{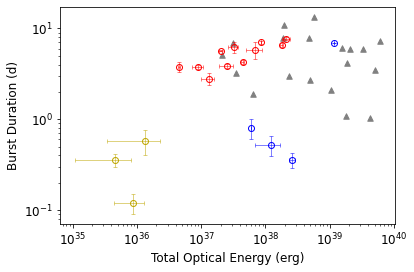}
    \end{subfigure}
    \vspace{0.1cm} 
    \begin{subfigure}[t]{0.48\textwidth}
        \centering
        \includegraphics[width=\linewidth]{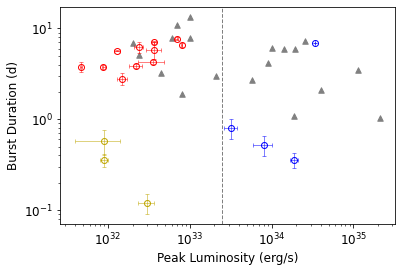}
    \end{subfigure}
    \hfill
    \begin{subfigure}[t]{0.48\textwidth}
        \centering
        \includegraphics[width=\linewidth]{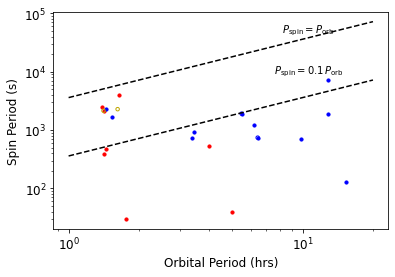}
    \end{subfigure}
    
    \caption{Properties of bursts in IPs similar to \citet{Ilkiewicz2024}. The top left, top right, and bottom left panels show the relationship between each of the properties measured and are all plotted on a logarithmic scale. Literature values for dwarf novae (red), micronovae (blue), and magnetic gating (yellow) are included with new additions from this research (grey) plotted simultaneously \citep{Ilkiewicz2024}. Note: Red dwarf nova points aren't all IPs. Bottom right: The spin-orbital period relationship for the new data with an arbitrary classification scheme defined in Section \ref{sec:BurstAnalysis} where the same colour scheme is applied. The grey dashed line on the top and bottom left panels indicates the classification criterion as described in Section \ref{sec:BurstComparison}.}
    \label{properties}
    \vspace{-0.1cm}
\end{figure*}

We reproduce the diagnostic diagrams of \citet{Ilkiewicz2024} using their data and methodology, then overlay our 19 newly identified bursts as gray triangles in Figure \ref{properties}. These data points are given as midpoints of the ranges in Table \ref{tab:outbursts}. Although our focus is on optical outbursts in intermediate polars, many of the dwarf novae reference points in these diagrams come from non‑magnetic systems, therefore caution should be exercised when making direct comparisons. For a complete list of those non‑magnetic systems and previously identified micronovae and magnetic gating bursts, see \citet{Ilkiewicz2024}. Furthermore, EI UMa, TV Col, and V1025 Cen are included in both data from \cite{Ilkiewicz2024} and this current research so these systems will appear twice on each plot in Figure \ref{properties}.

Since the uncertainties on our newly identified bursts are dominated by the lack of burst coverage due to low cadence observations, we do not plot individual error bars. Instead, these errors are characterized in our simulations (Section \ref{sec:ErrorSim}). In the top left panel of Figure \ref{properties}, many of the newly detected bursts appear to lie at higher energies than those that have previously been identified, while still following the same overall trend. It remains unclear whether dwarf novae and micronovae actually represent different populations here, however there does appear to be a shift in the overall trend at a luminosity of $\sim2.5\times10^{33}$erg s$^{-1}$. No magnetically gated bursts were detected above our significance threshold in our data set, however with these durations being below the \textit{ASAS-SN} and \textit{ZTF} cadence, this is not unexpected.

Assuming we have two different populations, we can define a preliminary classification scheme by setting a cut-off between dwarf novae and micronovae at $2.5\times10^{33}$erg s$^{-1}$, shown as a dashed grey line in the left panels of Figure \ref{properties}. Preliminarily, it appears we have identified 10 micronovae candidates and 9 dwarf novae candidates in IPs. Of these, 3 dwarf novae and 3 micronovae candidates match classifications from literature, while all of the other systems are entirely new additions or do not match their current literature classification.
In an attempt to understand how the type of burst is influenced by the evolutionary phase of the CV, where possible, we plot the spin to orbital period distribution for all of the bursts in Figure \ref{properties}. Spin and orbital periods are obtained directly from the IP catalogue webpage. This plot is seen on the bottom right panel of Figure \ref{properties} and shows an apparent distinction between the type of burst and the host CVs orbital period. Micronovae candidates appear to lie at longer orbital periods whereas dwarf novae candidates tend to concentrate in lower orbital period, more evolved binaries.

5 of the bursts in Table \ref{tab:outbursts} do not match their classification from literature. IGR J17014-4306 has been identified by \citet{Guerrero_2025} as a dwarf novae with low accretion rate. Outbursts in NY Lup have previously been reported by \citet{Hameury2022} and classified as dwarf novae. Outbursts in V1025 Cen and V1223 Oph were reported by \citet{Littlefield2022} and \citet{Hameury2022} but were classed as magnetic gating bursts, disagreeing with the burst energetics calculated in this work. V2731 Oph has recently been reported to have stunted outbursts, typically seen in nova-like variables which does not coincide with our micronovae classification \citep{Honeycutt_2025}. Note that 3 of the systems identified have orbital periods longer than 10 hours, the cut-off time used by \citet{Ilkiewicz2024} when developing diagnostic diagrams.

\section{Simulations}

In Section \ref{sec:ErrorSim} we discuss results of Monte Carlo simulations of the spread of peak luminosity and total energy as a result of measurement uncertainties and limited cadence. In Section \ref{sec:DetectionRate}, we apply a similar Monte Carlo process to evaluate how many micrnovae we would expect to miss in the 10 year observing period.

\subsection{Error Simulations}
\label{sec:ErrorSim}

Because ASAS-SN’s sparse sampling often misses the brightest phases of short-duration bursts, the estimated energetics, peak luminosities, and durations are systematically underestimated. Note that the energies and luminosities calculated here are already a lower limit since $\sim1/3-1/2$ of the burst energy is radiated in other parts of the spectrum e.g. X-rays. This limitation complicates efforts to compare events observed at higher cadence such as those with \textit{TESS} and assess their true physical origin. Rather than simply attaching formal error bars to individual measurements in Figure \ref{properties}, we performed dedicated Monte Carlo cadence-sampling simulations to quantify the impact of this bias on our results. These simulations provide a more realistic estimate of the uncertainties on the measurements displayed in Table \ref{tab:outbursts} as they incorporate the inconsistency of ASAS-SN sampling.

The Monte Carlo cadence-sampling simulation works as follows. We began with the full, 120s cadence \textit{TESS} light curve of the well‐characterized micronova, ASASSN-19bh \citep{Scaringi_2022b}. The choice of ASASSN-19bh as the outburst template does not affect the results, as the dispersion in peak luminosity and total energy shows little dependence on the shape of the profile. Using the typical \textit{ASAS-SN} g-band light curve of DQ Her as the baseline for \textit{ASAS-SN} cadence, we extracted the empirical distribution of time step intervals by computing the time differences between successive measurements, forming their cumulative distribution function (CDF). The CDF was then normalised between 0 and 1. For each iteration (1000 total) we generated a synthetic \textit{ASAS-SN} observing sequence by: 
\begin{enumerate}[label=(\roman*), leftmargin=*, widest=iii]
    \item Drawing a uniform random number $u\in[0,1]$.
    \item Mapping $u$ through the \textit{ASAS-SN} interval CDF to select a time step $\Delta t$.
    \item Adding $\Delta t$ to the previous synthetic epoch.
    \item Matching that epoch to the nearest time stamp within the \textit{TESS} light curve.
\end{enumerate}
For each iteration, the beginning of the synthetic \textit{ASAS-SN} light curve is randomised around 2337.6 days (BJD-2475000) with a range of $\pm3$ days. This process was repeated until the synthetic sequence extended across the full \textit{TESS} sector. On each iteration, the burst identification and photometric analysis pipeline was ran, computing the peak luminosity, total optical energy, and duration of each burst using the same method as in Section \ref{sec:BurstAnalysis}. Again, a threshold for burst detection as 3 consecutive points $3\sigma$ above the quiescent average is required for the pipeline to run.

\begin{figure*}
  \centering
    \includegraphics[width=\linewidth]{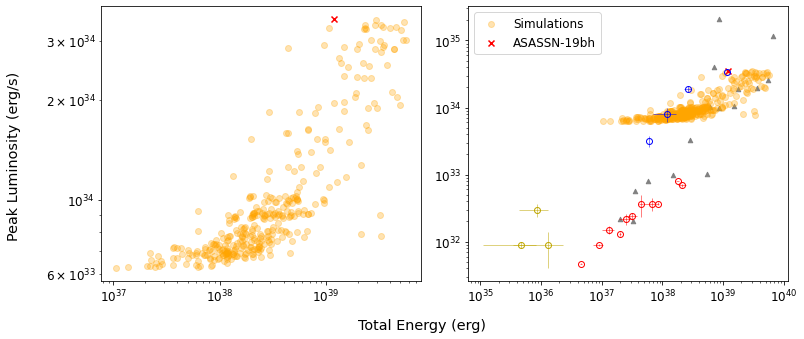}
  \caption{The results of monte-carlo simulations to quantify the error the method of calculating peak luminosity and total energy using ASASSN-19bh as a reference. The left panel displays the 1000 simulation points with their calculated peak luminosities and total energies in comparison with the position of the actual energy and luminosity of ASASSN-19bh obtained from \citet{Scaringi_2022b}. The right panel puts these results into context by overlaying simulations onto the top left panel of Figure \ref{properties}, where each point is represented by the same colour as in Figure \ref{properties}.}
  \label{fig:ErrorSim}
\end{figure*}

Figure \ref{fig:ErrorSim} shows the results of 1000 iterations of simulations. These simulations are shown in the context of the luminosity-energy diagram in the top left of Figure \ref{properties}. We see that compared to high-cadence surveys such as \textit{TESS}, we routinely underestimate luminosity and energy using our methodology. Notably we observe a range of 2 orders of magnitude in energy and almost one order of magnitude in luminosity. It is also worth noting that there is a small chance to overestimate energy also. A similar set of simulations has been performed to assess the accuracy of duration estimations and this is presented in Appendix \ref{sec:appendix}, Figure \ref{fig:append3}.

\subsection{Detection Rate Simulations}
\label{sec:DetectionRate}

A key limitation in observing micronovae lies not only in accurately determining properties, but in detection rate. Given the short timescales of micronovae, it is important to consider how the likelihood of their detection depends on observation time. We explored how often we would expect to miss micronovae altogether using low-cadence, all-sky surveys, such as \textit{ASAS-SN} or \textit{ZTF}, assuming bursts have a 1 year recurrence time using another set of Monte Carlo simulations. We note that the assumption of a 1 year recurrence time for micronovae is not well constrained and may vary between systems. Our results from this simulation would be weakened if true recurrence times were significantly different. To perform these simulations, we split micronovae into two types: Long-duration, such as that of ASASSN-19bh, and short-duration, as seen in TV Col \cite{Scaringi_2022b}. For each type, we extract a high–cadence \textit{TESS} light‐curve segment centred on the well–characterised micronova: BJD 2457000 + 2348–2355 (sector 38) for ASASSN-19bh, and 2197–2200 (sector 32) for TV Col. We then construct a conservative, long‐baseline synthetic \textit{TESS} light curve by repeating each segment once per year; a conservative estimate for the recurrence rate of micronovae. Between repetitions of the \textit{TESS} burst region, we generate a Gaussian scatter around the quiescent flux with standard deviation equal to that of the original light curve. Finally, to emulate survey sampling, we resample this extended time series onto the \textit{ASAS-SN} cadence for DQ Her by matching each synthetic \textit{TESS} timestamp to its nearest \textit{ASAS-SN} observation. DQ Her here is simply chosen as a typical \textit{ASAS-SN} light curve. 

To assess how survey length affects our ability to recover micronovae, we ran 25 sets of 500 simulations, where the effective observing timescale was stepped from 10 years down to 0.5 years. In every trial, we randomised the synthetic \textit{TESS} light curve by applying a uniform time offset of $\pm0-365.25$ days around an equilibrium value of 365.25 days. We then truncated the resulting time series to the target duration and resampled it onto the ASAS-SN cadence to compute the detection fraction as a function of survey length. An iteration is classed as a positive detection if three consecutive data points lie $3\sigma$ above the quiescent average. 

\begin{figure}
  \centering
    \includegraphics[width=\linewidth]{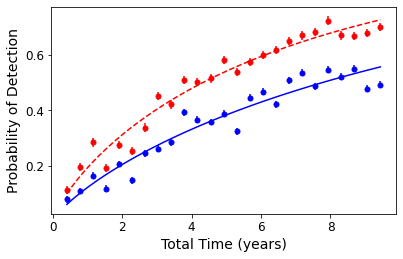}
  \caption{Probability of detection for 25 long term simulations, each with 500 iterations, with cut-off points between $\sim0.5$ and 10 years. The blue points and dotted line denote the data for ASASSN-19bh and its best fit using a saturating exponential function. The red points indicate data for TV Col.}
  \label{fig:DetRate}
\end{figure}

The proportion of synthetic light curves showing at least one burst, along with their Poisson errors, is plotted as a probability in Figure \ref{fig:DetRate} where the blue data represent ASASSN-19bh and the red data represent TV Col. Since we expect the probability to reach 1 as time tends towards infinity and that the probability should be 0 at the origin, we fit a curve with the function,
\begin{equation}
    P = 1-e^{-(\frac{t}{\tau})^k},
\end{equation}
where $P$ is the probability, $t$ is the length of observation, $k$ and $\tau$ are fitting parameters. For ASASSN-19bh, we find that $\tau=7.0\pm0.2$s and $k=0.76\pm0.3$ while for TV Col $\tau = 16\pm1$s and $k=0.71\pm0.04$.

We see that observations with a cadence similar to ASAS-SN take approximately 10 years to obtain a 70\% detection rate of micronovae (red line). However, this doesn't account for the fact that ASASSN-19bh, the micronova used for simulations, has an unusually long duration for the class. Typical micronovae, such as TV Col, have durations of $<$1 day \cite{Ilkiewicz2024}. We simulate that the detection rate drops significantly to $\sim$50\% after ten years in this more common class of micronovae (blue line).

\section{Discussion}
In Section \ref{sec:discussion} we discuss the results listed in Table \ref{tab:outbursts} and their reliability given the error simulations described in Section \ref{sec:ErrorSim}. Finally, in Section \ref{sec:DetectionRateDiscussion} we apply Monte Carlo simulations to test how many micronovae we expect to observe, for a given observing time, using \textit{ASAS-SN}.

\subsection{Outburst Population}
\label{sec:discussion}

As summarised in Table \ref{tab:outbursts}, we identify 20 confirmed or ironclad intermediate polars (IPs) exhibiting optical outbursts, corresponding to 28\% of all IPs. This is still far below the ratio of outbursting systems in non-magnetic  systems but represents an increase to the fraction reported by \citet{Hameury2017}. Six of these have been reported previously, and are marked with an asterisk in the table; the remainder represent new detections. Figure \ref{properties} presents four diagnostic diagrams that compare various burst properties which were previously thought to reveal clear distinctions between different classes of outburst events \citep{Ilkiewicz2024}. However, Figure \ref{properties} suggests that there is not as clear a distinction between dwarf novae and micronovae as was previously suggested. One possibility for this may lie in the fact that \citet{Ilkiewicz2024} considered a mix of magnetic and non-magnetic systems in their population analysis, while this work only considers IPs. It has been suggested by \cite{Hameury2017} that dwarf novae in IPs will be higher energy outbursts, potentially explaining some of the new detections we see here. However, the disc Instability Model (DIM) is unable to explain the shortest and highest energy outbursts observed in some systems, ruling out the possibility that they are dwarf novae \citep{Lasota_2001, Hameury2017}. Furthermore, \citet{Ilkiewicz2024} reported a dwarf nova observed in an IP with identical energetics to those seen in non-magnetic systems.

On the other hand, if we assume that the sample consists of two distinct populations, we might expect a separation between them at a luminosity of $\sim 2.5 \times 10^{33},\mathrm{erg,s^{-1}}$, as suggested by \citet{Ilkiewicz2024}. It is notable that the new points don't appear to follow the same separated linear trends above and below this threshold as those in \citet{Ilkiewicz2024}. However, we caution that our error simulations in Section \ref{sec:ErrorSim} suggest our methods may systematically underestimate these quantities. As a result, the potential scatter of the results in energy, luminosity, and duration is too large to obtain conclusions about the presence of two distinct populations. To further complicate this, some systems may exhibit both dwarf novae and micronovae. As a result, systems with multiple detected bursts may have their average properties contaminated by the presence of different classes of outburst. Future work could hope to classify individual outbursts with a higher cadence observations, comparable to \textit{TESS}.

Furthermore, \citet{Ilkiewicz2024} did not include CVs with an orbital period greater than 10 hours since they tend to have evolved donors which can exhibit flares of similar energy to micronovae \citep{Tu_2021}. The accretion discs in these systems are also larger leading to the possibility of higher energy dwarf novae outbursts being observed. However, in this work we have included a few such CVs: V2731 Oph, RX J2015.6+3711, and IGR J17014-4306. This is because, the analysis from Figure 9 of \citet{Tu_2021} suggests that the only bursts with energies similar to micronovae occur in giants. All donor stars within these systems have been suggested to have K/M type stars (V2731 Oph \citet{Lopes_de_Oliveira_2019}; RX J2015.6+3711 \citet{Coti_Zelati_2015}; IGR J17014–4306 (infer spectral K type from orbital period using \citet{Knigge2011} relation). The inclusion of these longer period CVs could explain why we see overlap between dwarf novae and micronovae populations as all of the above systems have been classified as micronovae. \citet{Ilkiewicz2024} excluded bursts with durations exceeding seven days to avoid contamination by more energetic dwarf novae. Several of the systems listed here (e.g. CC Scl and possibly HT Cam) reach or exceed that duration, potentially further blurring the distinction between dwarf novae and micronovae.

Again assuming that we have detected two separate populations of bursts and that classification is accurate, we see that dwarf novae tend to concentrate at lower orbital periods whereas micronovae are more prevalent in longer orbital period systems. This is an expected result since longer orbital period CVs, at an earlier stage in their evolution have higher mass transfer rates. Since the recurrence time for micronovae scales as $t_{rec}\propto \dot{M}^{-1}$ where $\dot{M}$ is the accretion rate onto the WD, we would expect micronovae to be more frequent in longer orbital period systems, therefore we have a much greater chance of observing them. \citet{Hameury2017} suggests that these short bursts could be attributed to an increased mass transfer rate from the secondary provided the quiescent mass transfer rate is low enough for the disc to remain stable. Since, most of these short bursts are detected at longer orbital periods, this appears unlikely.

Furthermore, we detect no magnetic gating systems in our data likely due to selection bias from our methods. Magnetic gating bursts tend to occur over a duration $\lesssim6$ hours \citep{Ilkiewicz2024}, and as such, we would likely only detect one data point using \textit{ASAS-SN}, whereas we require three consecutive points for a burst detection. Although micronovae are of only slightly longer duration, their higher energy and peak luminosity ensures a higher detection probability.

If we na\"ively assume that this arbitrary classification scheme is accurate and these are two separate populations, we have detected 10 micronovae candidates and 9 dwarf novae candidates within the IP population. With the systems not already reported in literature, this would more than double the known population of micronovae and provide more systems to study further and understand the origin of this phenomenon. Continued monitoring, coupled with rapid multi-wavelength follow-up during outbursts, will be essential for confirming their thermonuclear nature. A larger population would also enable statistical studies of how burst properties, such as duration and peak luminosity, are related. Such correlations may ultimately open up the possibility of using these events as standard candles \citep{Ilkiewicz2024}. 

As discussed in Section \ref{sec:ErrorSim}, the reliability of our findings depend on both the accuracy of the methods and data used. Since our simulations use micronovae as a sample, we likely observe an upper limit to the intrinsic scatter, directed by the extremely short duration of these bursts. As a result, we would expect micronovae to be more scattered around their intrinsic distribution than dwarf novae. The rare overestimates in energy are a result of extremely low simulated cadence, where trapezoidal integration does not recover the shape of the outburst accurately. It follows that we see the majority of overestimates in energy when the peak luminosity is closer to the true value. Ultimately, the simulated properties, as seen in context on the right panel of Figure \ref{fig:ErrorSim}, suggest that the bursts we detect are likely to have even higher intrinsic energies and luminosities. This would provide further support to the conclusion that micronovae are a separate population of outburst. On the other hand, simulations have also shown that some of these new detections can be overestimates, and would therefore more closely follow the linear trend identified by \citet{Ilkiewicz2024} previously. Although we cannot make specific conclusions about the prevalence of micronovae and dwarf novae in this sample, the detection of these bursts provide opportunity for future monitoring to characterise these bursts with a higher degree of accuracy. 

\subsection{Detection Rate Simulations}
\label{sec:DetectionRateDiscussion}
The simulations presented in Section \ref{sec:DetectionRate} indicate an expected detection probability of $\sim$50–70\% under the assumption of a 1-year recurrence time for micronovae. Consequently, we would expect archival \textit{ASAS-SN} data to reveal micronova in $\sim$50–70\% of IPs, whereas we observe only 14\%. This discrepancy may arise if the true recurrence timescale is longer than one year, which would increase the observing time needed to obtain the same statistics. Furthermore, our classifications remain preliminary and the significant scatter in the diagnostic diagrams, as described in Section \ref{sec:ErrorSim}, prevents us from confidently concluding anything dependent on the accuracy of these classifications. However, if we na\"ively assume that our classifications and recurrence time are accurate, then we could conclude that the entire population of IPs most likely doesn't display micronovae. We note, however, that this conclusion implicitly assumes that our detection algorithm, described in Section \ref{sec:BIA}, recovers all micronovae present in the data. This has been shown not to be the case, with 3 of the micronovae being detected by eye. In addition to observational biases, there are also physical mechanisms that may prevent micronovae from occurring in all IPs. \citet{Scaringi2022} explains that micronovae will be inhibited if the settling time of the column is faster than material accreted through $\dot{M}_{acc}$. Furthermore, a combination of mass transfer rate, surface magnetic field, WD spin, and spin-to-orbit alignment can provide unfavourable conditions to achieve $P_{\mathrm{base}} = P_{\mathrm{crit}}$, the requirement to trigger a micronovae \citep{Scaringi2022}. Here, $P_{\mathrm{base}}$ refers to the pressure at the base of the WD accretion column and $P_{\mathrm{crit}}$ is the critical pressure required to begin a thermonuclear reaction ($\sim10^{18}$ dyn cm$^{-2}$). Therefore, it is clear that there are many external factors that can inhibit micronovae and they likely would contribute to this discrepancy. Ultimately, these simulations should be used with caution and further high cadence observations of micronovae candidates are required to confirm this.

\section{Conclusions}

To conclude, we have detected 20 optical outbursts in ironclad IPs using archival \textit{ASAS-SN} data in combination with \textit{TESS} and \textit{ZTF}, out of which 16 haven't been previously reported in literature or were given a different classification. This corresponds to 28\% of all known IPs (as of 2021) and if we apply an arbitrary classification scheme, we find that similar proportions correspond to dwarf novae and micronovae, although this distinction remains uncertain due to large uncertainties. We detect no magnetic gating bursts, likely due to observational and search bias. We have organised bursts based on their peak luminosity, total energy and duration, finding that populations of dwarf novae and micronovae may appear to split at a luminosity of $\sim2.5\times10^{33}$ erg s$^{-1}$. Based on the diagnostic diagrams developed by \citet{Ilkiewicz2024}, it is unclear whether dwarf novae and micronvae classifications of the observed bursts result from different physical mechanisms however with higher cadence monitoring and more accurate burst energetics, a split between the two populations could become more apparent. Error simulations suggests that these results should be treated with caution as we systematically underestimate energy, duration and luminosity due to the low cadence of \textit{ASAS-SN}. Therefore, even though we cannot explicitly conclude that we have detected different populations of outbursts in this study, we provide a list of systems to target for future monitoring. Finally, according to detection rate simulations, we should expect to see between 60\% and 70\% of IPs displaying micronovae however this does not agree with the data. Therefore, if the assumptions underlying these simulations are accurate, we conclude that even though they are more widespread than previously understood, not all IPs show micronovae.

\section*{Acknowledgements}

This paper uses data collected from the ASAS-SN ground array of telescopes, where data has been downloaded directly from the ASAS-SN SkyPatrol V1.0 portal. Funding for the ASAS-SN team is provided by Gordon and Betty Moore foundation grants GBMF5490 and GBMF10501, Alfred P. Sloan Foundation 5-year grant and NSF grant AST-1908570 with telescopes being hosted by Las Cumbres University. This paper also includes data from the TESS mission obtained from the MAST data archive at the Space Telescope Science Institute (STScI). The TESS mission is funded by the NASA explorer program. STScI is operated by the Association of Universities for Research in Astronomy, Inc., under NASA contract NAS5–26555.  S.S. acknowladges STFC funding ST/T000244/1 and ST/X001075/1. M.V. acknowledges the support of the Science and Technologies Facilities Council (STFC) studentship ST/W507428/1. K.I. was supported by the Polish National Science Centre (NCN) grant 2024/55/D/ST9/01713

\section*{Data Availability}

ASAS-SN data used in this work \citep{Shappee_2014,Kochanek_2017} are available on the ASAS-SN webpage \hyperlink{ASAS-SN}{https://asas-sn.osu.edu/}. TESS data is available on the MAST webpage \hyperlink{MAST}{https://mast.stsci.edu/portal/Mashup/Clients/Mast/Portal.html}. ZTF data is availabe from the NASA/IPAC Infrared Science Archive \hyperlink{ZTF}{https://irsa.ipac.caltech.edu/Missions/ztf.html}.



\bibliographystyle{mnras}
\bibliography{references} 




\appendix
\section{Appendix}
\label{sec:appendix}

\begin{figure*}
  \centering

  \begin{subfigure}[t]{0.32\linewidth}\centering
    \includegraphics[width=\linewidth]{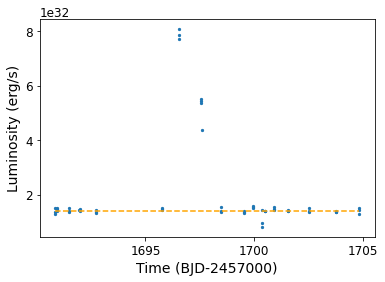}\caption{AX J1853.3-0128}
  \end{subfigure}\hfill
  \begin{subfigure}[t]{0.32\linewidth}\centering
    \includegraphics[width=\linewidth]{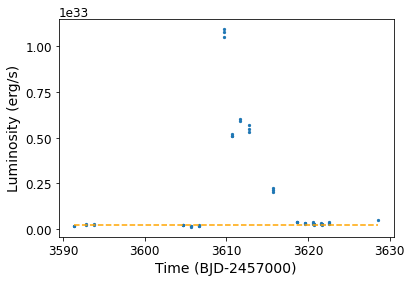}\caption{CC Scl}
  \end{subfigure}\hfill
  \begin{subfigure}[t]{0.32\linewidth}\centering
    \includegraphics[width=\linewidth]{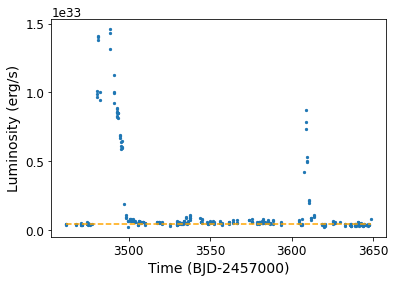}\caption{CTCV J2056-3014}
  \end{subfigure}

  \medskip

  \begin{subfigure}[t]{0.32\linewidth}\centering
    \includegraphics[width=\linewidth]{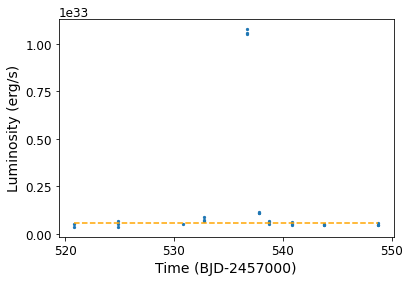}\caption{DO Dra}
  \end{subfigure}\hfill
  \begin{subfigure}[t]{0.32\linewidth}\centering
    \includegraphics[width=\linewidth]{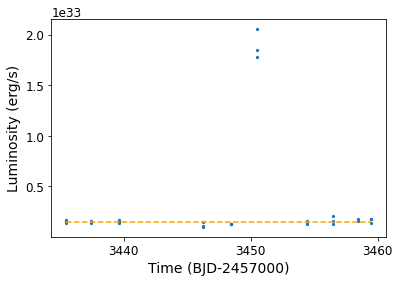}\caption{DW Cnc}
  \end{subfigure}\hfill
  \begin{subfigure}[t]{0.32\linewidth}\centering
    \includegraphics[width=\linewidth]{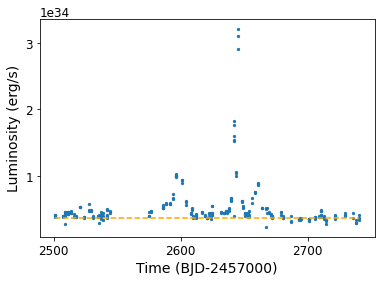}\caption{EI UMa}
  \end{subfigure}

  \medskip

  \begin{subfigure}[t]{0.32\linewidth}\centering
    \includegraphics[width=\linewidth]{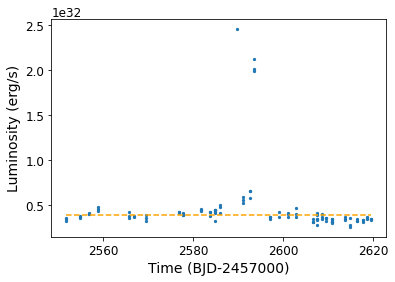}\caption{Ex Hya}
  \end{subfigure}\hfill
  \begin{subfigure}[t]{0.32\linewidth}\centering
    \includegraphics[width=\linewidth]{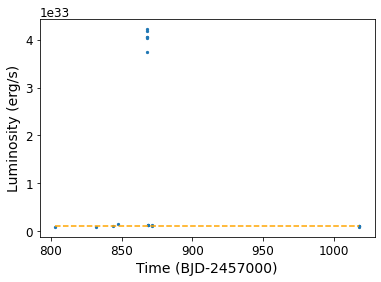}\caption{HT Cam}
  \end{subfigure}\hfill
  \begin{subfigure}[t]{0.32\linewidth}\centering
    \includegraphics[width=\linewidth]{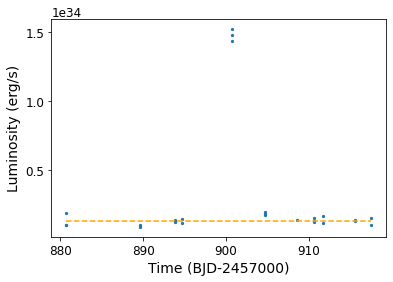}\caption{IGR J17014-4306}
  \end{subfigure}

  \caption{Examples of a single detected burst or two closely spaced bursts using \textit{ASAS-SN} for 9 of the IPs listed in Table \ref{tab:outbursts}. All bursts are detected in the g-band unless explicitly stated otherwise in the legend of the plot. The orange line in all panels displays the calculated quiescence level for each system.}
  \label{fig:Append1}
\end{figure*}

\begin{figure*}
  \centering

  \begin{subfigure}[t]{0.32\linewidth}\centering
    \includegraphics[width=\linewidth]{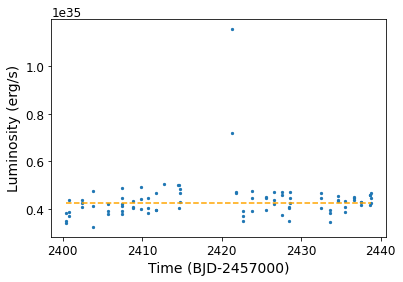}\caption{IGR J18173-2509}
  \end{subfigure}\hfill
  \begin{subfigure}[t]{0.32\linewidth}\centering
    \includegraphics[width=\linewidth]{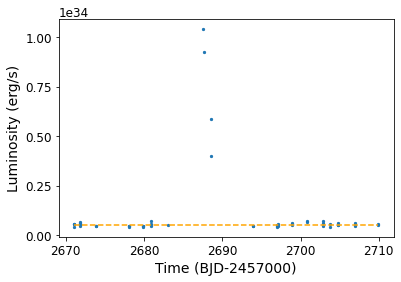}\caption{IGR J19267+1325}
  \end{subfigure}\hfill
  \begin{subfigure}[t]{0.32\linewidth}\centering
    \includegraphics[width=\linewidth]{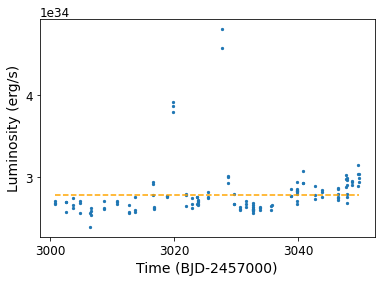}\caption{NY Lup}
  \end{subfigure}

  \medskip

  \begin{subfigure}[t]{0.32\linewidth}\centering
    \includegraphics[width=\linewidth]{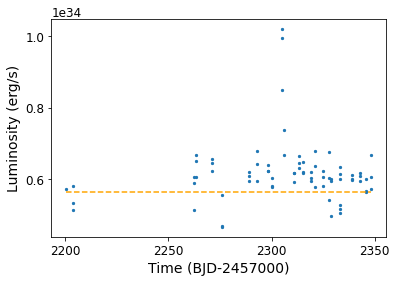}\caption{RX J2015.6+3711}
  \end{subfigure}\hfill
  \begin{subfigure}[t]{0.32\linewidth}\centering
    \includegraphics[width=\linewidth]{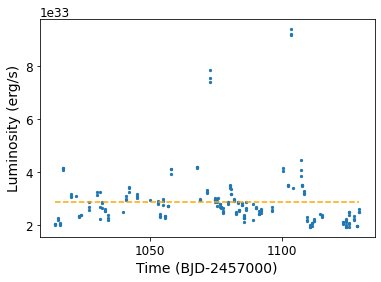}\caption{TV Col}
  \end{subfigure}\hfill
  \begin{subfigure}[t]{0.32\linewidth}\centering
    \includegraphics[width=\linewidth]{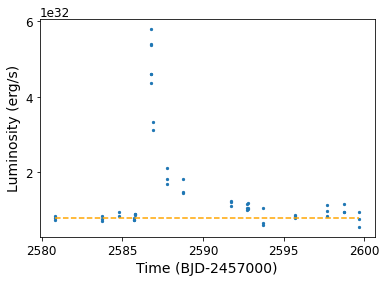}\caption{V1025 Cen}
  \end{subfigure}

  \medskip

  \begin{subfigure}[t]{0.32\linewidth}\centering
    \includegraphics[width=\linewidth]{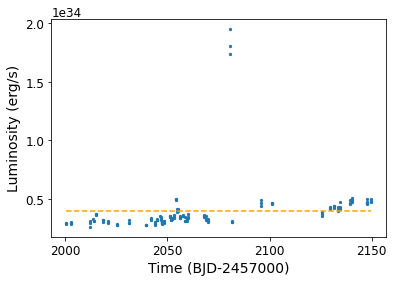}\caption{V1223 Sgr}
  \end{subfigure}\hfill
  \begin{subfigure}[t]{0.32\linewidth}\centering
    \includegraphics[width=\linewidth]{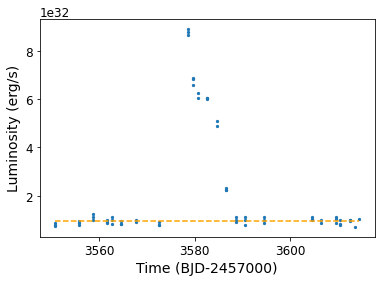}\caption{V1460 Her}
  \end{subfigure}\hfill
  \begin{subfigure}[t]{0.32\linewidth}\centering
    \includegraphics[width=\linewidth]{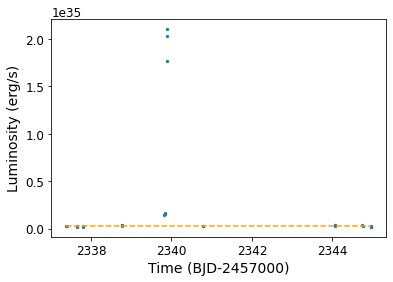}\caption{V2731 Oph}
  \end{subfigure}

  \medskip

  \begin{subfigure}[t]{0.32\linewidth}
    \centering
    \includegraphics[width=\linewidth]{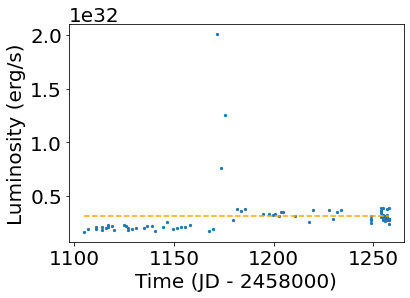}
    \caption{V598 Peg}
  \end{subfigure}
\hfill

  \caption{Examples of a single detected burst or multiple closely spaced bursts using \textit{ASAS-SN} for 10 of the IPs listed in Table \ref{tab:outbursts}. All bursts are detected in the g-band unless explicitly stated otherwise in the legend of the plot. The orange line in all panels displays the calculated quiescence level for each system.}
  \label{fig:Append2}
\end{figure*}

\begin{figure*}
  \centering
    \includegraphics[width=0.7\linewidth]{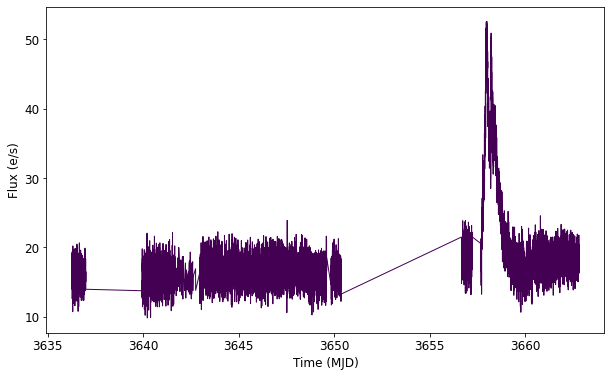}
  \caption{The \textit{TESS} lightcurve of IGR J04571+4527 in sector 86 showing a burst located at an MJD of 3658.}
  \label{fig:appendTESS}
\end{figure*}

Example bursts for each bursting IP are displayed in Figures \ref{fig:Append1} and \ref{fig:Append2}. Since these are only single-burst examples, they may not be representative of all bursts or the properties listed in Table \ref{tab:outbursts}. As explained in Section \ref{sec:BIA}, the error on the peak luminosity for each burst is given by a combination of the error on the distance to the source and the intrinsic error from the data. For sources with only a single burst detected, the error on the luminosity in erg s$^{-1}$ is propagated according to:
\begin{equation}
    \sigma_L = L[\mathrm{erg s}^{-1}] \times \sqrt{(2\frac{\sigma_d}{d})^2+(\frac{\sigma_F[Jy]}{F[Jy]})^2},
\label{eq:err_single}
\end{equation}
where $\sigma_L$ is the error on the luminosity, $\sigma_d$ is the error on the distance $d$, and $F[Jy]$ is the luminosity directly from \textit{ASAS-SN} which has an intrinsic error $\sigma_F[Jy]$. The presence of multiple bursts complicates this as the dispersion of the peak luminosities must be taken into account. Therefore the final error on sources with multiple bursts is given by:
\begin{equation}
\sigma_{\bar{L}} =
\sqrt{
\frac{1}{N^2} \sum_{i=1}^{N} \sigma_{F}^2
\;+\;
\frac{s^2}{N}
\;+\;
\sigma_{D}^2
},
\label{eq:err_multi}
\end{equation}
where $s$ is the standard deviation of the peak luminosities, and $N$ is the number of bursts used in the calculation of the mean. $\sigma_D$ is the total error on the distance and is given by $\sigma_D = L\times\frac{2\sigma_d}{d}$. In most cases, the spread in peak luminosities is the dominant source of error.

\begin{figure*}
  \centering
    \includegraphics[width=\linewidth]{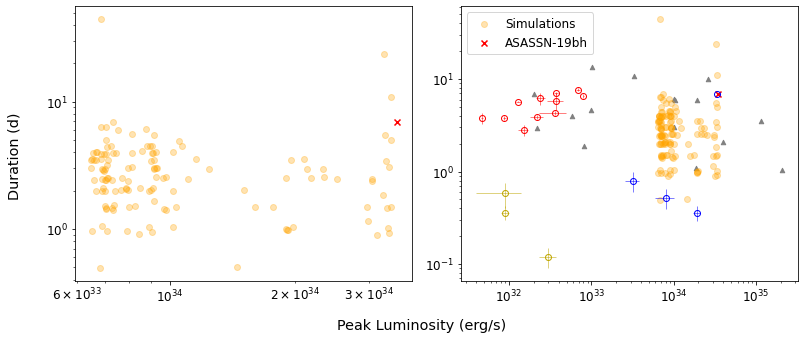}
  \caption{The results of monte-carlo simulations to quantify the error the method of calculating peak luminosity and duration. The left panel displays the 400 simulation points with their calculated peak luminosities and total energies in comparison with the position of the actual energy and luminosity of ASASSN-19bh obtained from \citet{Scaringi_2022b}. The right panel puts these results into context by overlaying simulations onto the bottom left panel of Figure \ref{properties}, where each point is represented by the same colour as in Figure \ref{properties}.}
  \label{fig:append3}
\end{figure*}

Using the same process as described in Section \ref{sec:ErrorSim}, we simulated the systematic effects of data cadence and methodology on our measurement of duration. Similarly to energy, we are most likely to underestimate duration however there is a small probability of overestimation as shown in Figure \ref{fig:append3}. Figure \ref{fig:append3} also shows these results in the context of the bottom left panel of Figure \ref{properties}.

\begin{figure*}
  \centering
    \includegraphics[width=300pt]{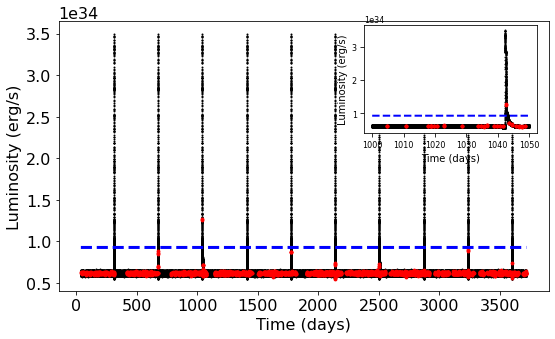}
  \caption{One iteration of a 10-year ASAS-SN simulation. Black points show a synthetic repeating light curve of ASASSN-19bh using \textit{TESS} sector 38 data. Red points represent the simulated ASAS-SN cadence, and the blue dashed line marks the burst detection threshold. The inset zooms in on a detected burst.}
  \label{fig:append4}
\end{figure*}

Finally, an example of a single iteration from the detection rate simulation is shown in Figure \ref{fig:append4}. This displays a single iteration at 10 years, with micronovae repeating once per year. The red points represent the results of one simulation iteration with the inset plot visualising a burst detection above threshold.

\bsp	
\label{lastpage}
\end{document}